\documentclass{biophys-new}
\usepackage[utf8]{inputenc}
\usepackage{graphicx}
\usepackage[colorlinks,allcolors=cyan!70!black]{hyperref}

\usepackage{lipsum}

\title{A Mechanical Instability in Planar Epithelial Monolayers Leads to Cell Extrusion}
\runningtitle{Mechanics of Epithelial Extrusion} 

\author[1,2,*]{Satoru Okuda}
\author[3]{Koichi Fujimoto}
\runningauthor{Okuda and Fujimoto} 

\affil[1]{WPI Nano Life Science Institute (WPI-NanoLSI), Kanazawa University, Kakuma-machi, Kanazawa 920-1192, Japan}
\affil[2]{PRESTO, Japan Science and Technology Agency, Kyoto 606-8507, Japan}
\affil[3]{Graduate School of Sciences, Osaka University, Toyonaka 560-0043, Japan}

\corrauthor[*]{satokuda@staff.kanazawa-u.ac.jp}

\papertype{Article}

\begin{document}

\begin{frontmatter}

\begin{abstract} 
In cell extrusion, a cell embedded in an epithelial monolayer loses its apical or basal surface and is subsequently squeezed out of the monolayer by neighboring cells. Cell extrusions occur during apoptosis, epithelial-mesenchymal transition, or pre-cancerous cell invasion. They play important roles in embryogenesis, homeostasis, carcinogenesis, and many other biological processes. Although many of the molecular factors involved in cell extrusion are known, little is known about the mechanical basis of cell extrusion. We used a three-dimensional (3D) vertex model to investigate the mechanical stability of cells arranged in a monolayer with 3D foam geometry. We found that when the cells composing the monolayer have homogeneous mechanical properties, cells are extruded from the monolayer when the symmetry of the 3D geometry is broken due to an increase in cell density or a decrease in the number of topological neighbors around single cells. Those results suggest that mechanical instability inherent in the 3D foam geometry of epithelial monolayers is sufficient to drive epithelial cell extrusion. In the situation where cells in the monolayer actively generate contractile or adhesive forces under the control of intrinsic genetic programs, the forces act to break the symmetry of the monolayer, leading to cell extrusion that is directed to the apical or basal side of the monolayer by the balance of contractile and adhesive forces on the apical and basal sides. Although our analyses are based on a simple mechanical model, our results are in accordance with observations of epithelial monolayers {\it in vivo} and consistently explain cell extrusions under a wide range of physiological and pathophysiological conditions. Our results illustrate the importance of a mechanical understanding of cell extrusion and provide a basis by which to link molecular regulation to physical processes.
\end{abstract}

\end{frontmatter}

\section*{Statement of Significance}
Epithelial cell extrusion is important for biological processes such as embryogenesis, homeostasis, and carcinogenesis. Various molecular factors, such as cancer genes and their products, have been reported as key drivers of epithelial extrusion; however, little is known about how these factors are linked to the mechanical process. A simple mathematical model based on mechanics can consistently explain cell extrusions under a wide range of physiological and pathophysiological conditions. The model shows that cells can be extruded from homogeneous sheets, owing to the inherent mechanical instability of the 3D foam geometry of the epithelial monolayer. When the cells generate active forces, the forces act to enhance the instability and direct extrusion to either the apical or basal side of the monolayer.

\section*{INTRODUCTION}
Foam geometry is ubiquitous in nature, appearing in contexts ranging from the large-scale structure of the cosmos to the froth on a glass of beer. Epithelial sheets are an example of living tissues with foam geometry, referred to as ``cell packing geometry'' in biological terms. Epithelial sheets are monolayers of epithelial cells that have the ability to dynamically change their shape and three-dimensional (3D) configuration, as is widely observed in morphogenesis, homeostasis, and carcinogenesis \cite{slattum2014tumour,lee2006epithelial,thiery2009epithelial,macara2014epithelial}.
Epithelial cells usually possess both an apical surface and a basal surface, which help to maintain the integrity of the monolayer. Occasionally, a single cell loses its apical or basal surface and is extruded from the monolayer to the side opposite that of the lost surface. The process of epithelial cell extrusion is also referred to as delamination or protrusion. Examples of epithelial extrusion in vertebrates include the extrusion of apoptotic cells to the apical side of the monolayer as part of homeostasis and the extrusion of pre-cancerous cells to the basal side of the monolayer as part of tumor growth and carcinogenesis \cite{slattum2014tumour}. Extrusions to the basal side of the monolayer also occur in epithelial-mesenchymal transitions in vertebrates and invertebrates \cite{lee2006epithelial,thiery2009epithelial}. 


Epithelial extrusion is a mechanical process of the epithelial monolayer. Upon extrusion, the epithelial structure transits from a
symmetric monolayer to a multilayer that is asymmetric relative to the apicobasal axis.
Although many studies have focused on the molecular mechanisms of cell extrusions in various physiological contexts, an understanding of the mechanical basis of cell extrusion is still lacking. Recent studies showed that mechanical factors play key roles in the regulation of cell extrusion, including actomyosin contractility and cadherin or integrin-mediated adhesion \cite{katoh2012epithelial,gudipaty2017epithelial,rosenblatt2001epithelial,hogan2009characterization,monier2015apico,kajita2014filamin,friedl2003tumour}, cell crowding \cite{marinari2012live,eisenhoffer2012crowding,levayer2016tissue}, and the force balance on the apical junctional network \cite{marinari2012live,tsuboi2018competition,ohsawa2018cell}. Some of the mechanical factors that regulate cell extrusion are regulated by genetic programs; however, others are based on the geometry and stability of the cellular arrangement. During extrusion, multiple mechanical forces driven by disparate underlying phenomena act together in 3D space. In order to gain a consistent understanding of cell extrusion, it is necessary to clarify the contribution of each mechanical force within the 3D multicellular geometry. 

The physical approach to understanding cell extrusion is gradually gathering attention. For example, Saw et al. modeled the epithelium as an active nematic liquid crystal and suggested that apoptotic cell extrusions are driven by topological defects of cellular alignments \cite{saw2017topological}.
Remarkable progress has been made in the development of mechanical descriptions of epithelial cell geometries \cite{hannezo2014theory,honda1982cell,farhadifar2007the,sedzinski2016emergence,misra2016shape}. Pioneering work by Hannezo et al. explained buckling instabilities of the epithelial sheet on the basis of the 3D force balance among apical, lateral, and basal components \cite{hannezo2014theory}. 
That and other work used a mean-field approximation of cell shapes \cite{hannezo2014theory,honda1982cell,farhadifar2007the,misra2016shape},
 which implicitly assumes that cells robustly maintain a homogeneous monolayer sheet structure. On the other hand, Hannezo et al. explained how a cell embedded inside a tissue is moved to the surface of the tissue by using an in-plane approximation to consider the dynamics of the apical shapes of single cells \cite{sedzinski2016emergence}. 
Epithelial cell extrusion occurs locally in epithelial sheets with apicobasal asymmetry \cite{katoh2012epithelial,gudipaty2017epithelial,rosenblatt2001epithelial,hogan2009characterization,monier2015apico,kajita2014filamin}. Therefore, a discrete and 3D description of multicellular mechanics (rather than a mean-field or in-plane description) is needed to elucidate its mechanical basis.

In this study, we present a description of epithelial extrusion using a 3D vertex model \cite{honda2004a,okuda2013reversible,hashimoto2018topological}. First, we consider the situation where an extruding cell and its neighbors possess homogeneous mechanical properties. We analytically calculate energy landscapes and show that mechanical instability leading to cell extrusion is inherent in the 3D foam geometry of the epithelial monolayer. Our results also clarify the individual effects of mechanical properties, cell density, and topology on cell extrusion. Second, we consider the situation where active forces are generated on the apical, lateral, or basal region of an extruding cell according to intrinsic genetic programs. Our results show that the active forces provoke the inherent mechanical instability of the monolayer to initiate the extrusion process as well as direct the extrusion to the apical or basal side of the monolayer.

\section*{MATERIALS AND METHODS}

\subsection*{A model of the 3D foam geometry of epithelial cell sheets}

\begin{figure}
\begin{center}
\includegraphics[scale=1.5]{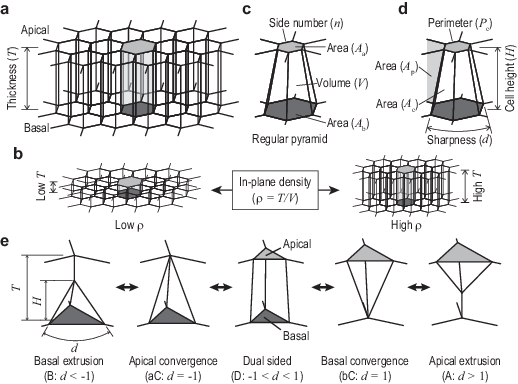}
\end{center}
\caption{
Mathematical model of the 3D foam geometry of epithelial sheets.
{\bf a}. Geometry of a cell sheet with thickness $T$. 
{\bf b}. In-plane cell density $\rho$ introduced to replace $T$. $\rho$ is defined as $T / V$, where $V$ is the average cell volume.
{\bf c}. Geometry of a single center cell embedded in the sheet. The geometry is defined as a regular frustum, whose shape is uniquely determined by the number of topological neighbors $n$, volume $V$, the apical surface area $A_\text{a}$, and the basal surface area $A_\text{b}$ of the center cell.
{\bf d}. Geometric parameters of the center cell: $H$, height of the center cell; $d$, degree of sharpness defined by Eq. \ref{sharpness}; $P_\text{c}$, apical perimeter of the center cell; $A_\text{c}$, total area of $n$ boundary faces between the center cell and its first neighbors; $A_\text{p}$, total area of $n$ boundary faces between first-neighbor cells. 
{\bf e}. Physical states and topology of cells embedded in the planar sheet characterized by sharpness $d$.
}
\label{fig:1}
\end{figure}

We created a geometric model of a cell sheet in which the cells are represented as polyhedrons with average volume $V$ ($>0$) within a planar sheet with homogeneous thickness $T$ ($>0$) (Fig. \ref{fig:1}a). We parameterized $T$ by introducing cell density into the sheet $\rho$ ($>0$) (Fig. \ref{fig:1}b). The effective area of each individual cell in the sheet is given by $V/T$, so the in-plane cell density is given as $\rho = T/V$.
The in-plane cell density corresponds to the aspect ratio of the cells, which can be columnar ($\rho \gg V^{-\frac{2}{3}}$), cuboidal ($\rho \approx V^{-\frac{2}{3}}$), or squamous ($\rho \ll V^{-\frac{2}{3}}$) in shape. For simplification, we assumed that the apical and basal surfaces are constrained in the plane and considered only the movements of a single center cell and its first neighbors within the sheet (Fig. \ref{fig:1}c,d), while keeping the other neighboring cells fixed in position. Under that approximation, the in-plane cell density $\rho$ implicitly reflects the effects of the surrounding cells (i.e., the mechanical environment).

We modeled the center cell as an $n$-sided regular frustum ($n \ge 3$) with volume $V$, apical surface area $A_\text{a}$ ($\ge 0$), and basal surface area $A_\text{b}$ ($\ge 0$) (Fig. \ref{fig:1}c).
The set of $n$, $V$, $A_\text{a}$, and $A_\text{b}$ uniquely determines the shape of the center cell. 
In that geometry, the center cell has either an apical surface, a basal surface, or both and is adjacent to $n$  first-neighbor cells with which it shares lateral boundary faces. The $n$ first-neighbor cells also lateral boundary faces with one another and are aligned radially around the center cell. Additionally, we introduced several geometric parameters (Fig. \ref{fig:1}d) including the height of the center cell $H$ ($0 < h \le T$), the apical perimeter of the center cell $P_\text{c}$ ($\ge 0$), the total area of the $n$ boundary faces between the center cell and its first neighboring cells $A_\text{c}$ $(> 0$), and total area of the $n$ boundary faces between the first-neighbors cells $A_\text{p}$. 

A key part of our model is cell rearrangements in the out-of-plane direction (Fig. \ref{fig:1}e), in which the surface area of the center cell shrinks to zero at either the apical surface ($A_\text{a}=0$) or the basal surface ($A_\text{b}=0$) and the cell is then extruded from the side of the monolayer opposite that of the lost surface. Such rearrangements are expressed by the degree of sharpness $d$ (Fig. \ref{fig:1}d,e), which is defined as
\begin{equation}
\label{sharpness}
d = \frac{T}{H} \frac{ \sqrt{A_\text{a}} - \sqrt{A_\text{b}}}{\sqrt{A_\text{a}}+\sqrt{A_\text{b}}}.
\end{equation}
The parameter $d$ characterizes the cell states continuously over topological rearrangements of the 3D cell configuration (Fig. \ref{fig:1}e). Thus, the rearrangements can be classified as basal extrusion (state B: $d<-1$), apical convergence (state aC: $d=-1$), dual-sided (state D: $-1<d<1$), basal convergence (state bC: $d=1$), or apical extrusion (state A: $d>1$).
In the apical/basal convergence states (aC and bC), the cells locally form pseudostratified structures, as observed in tissues such as neuroepithelia \cite{miyata2001asymmetric,haubensak2004neurons} and bronchial epithelia \cite{rackley2012building}.

\subsection*{The mechanical energy of epithelial sheets as a function of 3D geometric parameters}

\begin{figure}
\begin{center}
\includegraphics[scale=1.5]{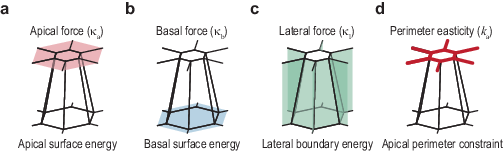}
\end{center}
\caption{
Mechanical forces exerted on the center cell and its first neighbors, as described by Eq. \ref{TotalEnergy}. 
{\bf a}. Apical surface tension exerted on the apical area of each cell. {\bf b}. Basal surface energy exerted on the basal area of each cell. {\bf c}. Lateral boundary energy exerted on the boundary areas among the cells. {\bf d}. Apical perimeter constraint exerted on the apical perimeter of each cell.
}
\label{fig:2}
\end{figure}

We model the mechanical energy of the epithelial sheet as follows. First, we prescribed a fixed volume for each cell, because the cell volume has been shown to be almost constant during morphological changes in several organisms \cite{weber2007tmod3,gelbart2012volume}.
Even under the constraints of constant cell volume and planar-sheet geometry, the cell-cell boundaries can move in principle, so $d$ remains variable. Epithelial sheets typically have a cortical actomyosin meshwork lining the apical surfaces of the cells \cite{lecuit2007cell}, and actin stress fibers and integrin on the basal cell surfaces to anchor the cells to the substrate \cite{berrier2007cell}. The cortical force is also exerted on the lateral boundaries between cells \cite{okuda2013reversible,hannezo2014theory,bielmeier2016interface}. To model the forces generated by those molecules, we introduced tension and adhesion onto the individual cell surfaces, depending on the apicobasal polarity \cite{dawes2005folded}. Those surface energies are usually modeled as first-order approximations proportional to the individual surface areas \cite{okuda2013reversible,hannezo2014theory,bielmeier2016interface}. We therefore made the energies of the $i$th apical and basal surfaces proportional to the $i$th apical and basal surface areas (i.e.,$\kappa_\text{a} A_{\text{a}}^i$ and $\kappa_\text{b} A_{\text{b}}^i$, respectively, where $\kappa_\text{a}$ and $\kappa_\text{b}$ are positive interfacial tensions; Fig.\ref{fig:2}a,b). Similarly , we made the energy of the boundary between the $i$th and $j$th cells proportional to the boundary area between those cells (i.e.,$\kappa_\text{l} A^{ij}$, where $\kappa_\text{l}$ is the interfacial tension; Fig.\ref{fig:2}c). Finally, we introduced contractile and adhesive forces along the apical perimeters of the cells to provide a force balance that acts to maintain a preferred apical perimeter [\cite{lecuit2007cell, farhadifar2007the}; Fig. \ref{fig:2}d]. Specifically, we modeled the energy of the $i$th apical perimeter $P^{i}$ as an elastic energy around a preferred perimeter value $P_\text{eq}$ [i.e., $k_\text{a} \left( P^{i} - P_\text{eq} \right)^2$, where $k_\text{a}$ is a non-negative elastic modulus \cite{farhadifar2007the,misra2016shape}]. Assuming that the cells are arranged in a planar-sheet geometry, we set $P_\text{eq}$ so that the cells would take the shape of a straight prism. The mechanical energy of the entire epithelial cell sheet is given as:

\begin{equation}
\label{TotalEnergy}
U_\text{e} = 
  \sum^\text{cells}_i \kappa_\text{a} A_{\text{a}}^i
+ \sum^\text{cells}_i \kappa_\text{b} A_{\text{b}}^i
+ \sum^\text{cells}_{ij} \kappa_\text{l} A^{ij}
+ \sum^\text{cells}_i k_\text{a} \left( P^{i} - P_\text{eq} \right)^2.
\end{equation}


The second and third terms in Eq. \ref{TotalEnergy} are negligible, because the total area of each apical and basal surface is conserved across the entire sheet. On the other hand, the fourth and last terms are effective; that is, because the positions of the second and nearest-neighbor cells are fixed, we expanded the fourth term around the center cell and first-neighbor cells as $\kappa_\text{l} \left(A_\text{c} + A_\text{p} \right)$ (Fig. \ref{fig:1}d). Similarly, we expanded the last term as $k_\text{a} \{ 1 + (1/n) [ 1 - \csc ( \pi / n ) ) ]^2 \} ( P_\text{c} - P_\text{eq} )^2$, where $\{ 1 + (1/n) [ 1 - \csc ( \pi / n ) ]^2 \}$ is a correction factor that  reflects the effects of the center cell and first-neighbor cells (Eq. \ref{surroundingperimeter} in the Supplementary Material).
Thus, Eq. \ref{TotalEnergy} can be rewritten as the mechanical energy of the center cell and first-neighbor cells in the planar sheet:
\begin{equation}
\label{LocalEnergy}
U_\text{p} = 
  k_\text{a} \left\{ 1 + \cfrac{1}{n} \left[ 1 -\csc \left( \frac{\pi}{n} \right) \right]^2 \right\} \left( P_\text{c} - P_\text{eq} \right)^2
+ \kappa_\text{l} \left(A_\text{c} + A_\text{p} \right).
\end{equation}

The packing geometry imposes the geometric constraints on the physical parameters (i.e., Eqs. \ref{frustum} and \ref{topology} in the Supplementary Material). We therefore solved Eqs. \ref{sharpness}, \ref{frustum} and \ref{topology} simultaneously and replaced $T$ with $\rho$, using $\rho=T/V$. The calculation gave $H$, $A_\text{a}$, and $A_\text{b}$ as functions of $\rho$, $V$, and $d$ (Eqs. \ref{cellheight}, \ref{apicalarea}, and \ref{basalarea} in the Supplementary Material). Thus, the set of parameters that uniquely determines the shape of the center cell was replaced with $n$, $\rho$, $V$, and $d$. In addition, the geometric parameters $P_\text{c}$, $A_\text{c}$, and $A_\text{p}$ in the energy function $U$ were expressed as functions of $n$, $\rho$, $V$, and $d$ (Eqs. \ref{apicalperimeter2}, \ref{lateralarea2}, and \ref{surroundingarea2} in the Supplementary Material), giving $U = U \left( n, \rho, V, d \right)$. 
Using the analytical expressions of $U$, we then calculated energy landscapes as functions of $d$ under specific values of $n$, $\rho$, and $V$. Similarly, we also obtained state diagrams by calculating the value of $d$ that minimizes $U$ under arbitrary values of $n$, $\rho$, and $V$.

\section*{RESULTS}

\subsection*{Mechanical instability is inherent in the 3D foam geometry of the cell sheet}

Cell extrusions occasionally occur even without the active generation of  force by the extruding cells, suggesting that mechanical instability inherent in the 3D geometry of the epithelial monolayer might be sufficient to drive cell extrusion. To analyze the mechanical stability of a theoretical epithelial monolayer, we focused on the behavior of Eq. \ref{LocalEnergy} around $d=0$.

Under physiological conditions, extruding cells tend to lose their apicobasal polarity, and the adherens junction and apical circumferential belt tend to disappear. The apical constraint ($k_\text{a}$) expresses the summation of the mechanical forces generated by the adherens junction and the apical circumferential belt.

By assuming $k_\text{a}=0$, we obtained the McLaurin expansion of Eq. \ref{LocalEnergy} for $d$ as
\begin{equation}
\label{McLaurin}
U_\text{p} = \begin{array}{ll}
  f_\text{0} \kappa_\text{l} 
+ f_\text{2} \kappa_\text{l} d^2 
+ O \left( d^4 \right)
& \text{as $-1 < d < 1$}
\end{array}
,
\end{equation}
where $f_\text{0}$ and $f_\text{2}$ are the zeroth and second-order coefficients of $d$ as functions of $n$, $\rho$, and $V$. Equation \ref{McLaurin} is an even function, because Eq. \ref{LocalEnergy} is symmetric about $d=0$ under $k_\text{a}=0$. The state at $d=0$ is stable when $f_\text{2} > 0$ and becomes unstable when $f_\text{2} < 0$; that is, state bifurcations occur in a symmetry-breaking manner, depending on $n$, $\rho$, and $V$. The dependence on $f_\text{2}$ is transformed to dependence on $\rho$ by the analytical expression of $f_\text{2}$ (Eq. \ref{McLaurin2} in the Supplementary Material). Thus, the state at $d=0$ is stable when $\rho < \rho_\text{*}$ and unstable when $\rho > \rho_\text{*}$, where $\rho_\text{*}$ is the critical density described as 
\begin{equation}
\label{Criticaldensity}
\rho_\text{*} = 2 \left( \frac{ 3 \cos \left( \frac{\pi}{n} \right) }{ n V^2 \left( 2 \sin \left( \frac{\pi}{n} \right) - 1 \right) } \right)^\frac{1}{3}.
\end{equation}
In Eq. \ref{Criticaldensity}, $\rho_\text{*}$ is positive when $n \le 6$ and negative when $n \ge 7$. In other words, the state at $d=0$ can be unstable when $n \le 6$, depending on $\rho$, whereas it is always stable and independent of $\rho$ when $n \ge 7$. Those results indicate an inherent mechanical instability in the 3D foam geometry of the cell-cell boundaries in the cellular sheet. Because of that inherent instability, mechanical disturbances in the form of changes in cell topology or cell density within the monolayer can induce extrusion without the application of any additional forces.


\subsection*{Cell density and topology regulate the mechanical instability within the epithelial sheet}

\begin{figure}
\begin{center}
\includegraphics[scale=1.5]{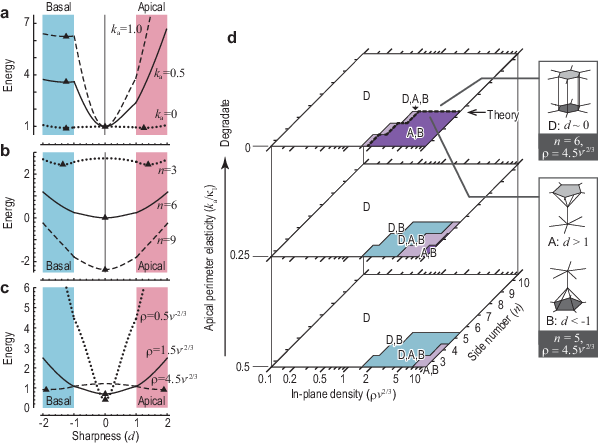}
\end{center}
\caption{
Mechanical stability of a homogeneous epithelial sheet with respect to apical elasticity, cell density, and topology.
{\bf a}--{\bf c}. Energy landscapes of Eq. \ref{LocalEnergy} with respect to the apical perimeter elasticity $k_\text{a}$, the number of sides $n$, and the in-plane cell density $\rho$.
Triangles ($\blacktriangle$) indicate energy minima.
Physical parameters were set to $n = 5$ and $\rho = 3 V^{-\frac{2}{3}}$ in {\bf a}, $k_\text{a} = 0$ and $\rho = 2.0 V^{-\frac{2}{3}}$ in {\bf b}, and $k_\text{a} = 0$ and $n = 5$ in {\bf c}.
In a--c, apical ($>1$) and basal ($<-1$) extrusion states are colored red and blue, respectively.
{\bf d}. State diagram of Eq. \ref{LocalEnergy} with respect to the number of sides $n$, the in-plane cell density $\rho$, and the apical perimeter constraint $k_\text{a}$. 
The cell states are color coded: white unistable state (D); purple, bistable states (A) and (B); light blue, bistable states (D) and (B); light purple, tristable states (D), (A), and (B). The states (D), (A), and (B) are identical to those in Fig. \ref{fig:1}e. The dashed line indicates the theoretical transition boundary between stable and unstable regions of dual-sided state (D), as described by Eq. \ref{Criticaldensity}.
}
\label{fig:3}
\end{figure}

To determine the effects of geometric and physical parameters on cell extrusion in more detail, we analyzed the energy landscape of Eq. \ref{LocalEnergy} for $k_\text{a}$, $n$, and $\rho$.

As expected from Eq. \ref{McLaurin}, the energy landscape of Eq. \ref{LocalEnergy} led to mechanical instability and subsequent extrusion (Fig. \ref{fig:3}). For example, the decrease in $k_\text{a}$ destabilized the dual-sided state (D) and stabilized the apical extrusion state (A) (Fig. \ref{fig:3}a). Both a decrease in $n$ and an increase in $\rho$ destabilized the dual-sided state (D) and stabilized the apical and basal extrusion states (A and B) (Fig. \ref{fig:3}b,c).
The state diagram (Fig. \ref{fig:3}d) indicates that the stability of the dual-sided state (D) and the apical extrusion state (A) depends on all three parameters $k_\text{a}$, $n$, and $\rho$. For example, the dual-sided state (D) can be destabilized by a decrease in $k_\text{a}$, a decrease in $n$, or an increase in $\rho$. On the other hand, the stability of basal extrusion state (B) depends on $n$ and $\rho$ but not on $k_\text{a}$. That is, the number of neighbors $n$ and the in-plane density $\rho$ affects all regions of $d$ (Fig. \ref{fig:3}b-c), whereas the apical perimeter elasticity $k_a$ affects only a specific region of $d$ ($>-1$) (Fig. \ref{fig:3}a). The difference in the state stabilities causes several bistable states (D and B) and tristable states (D, A, and B) (Fig. \ref{fig:3}d). From a quantitative point of view, the apical and/or extrusion states (A and B) can be stabilized for $n \le 5$ and $\rho \gtrsim 1.3 V^{-\frac{2}{3}}$ (Fig. \ref{fig:3}d). 
Specifically, the transition boundary between stable and unstable for the dual-sided state (D) obeyed the theoretical boundary of Eq. \ref{Criticaldensity} (dashed line at $k_\text{a}=0$ in Fig. \ref{fig:3}d). The dependence on $k_\text{a}$, $n$ and $\rho$ indicates that the 3D foam geometry of the cellular sheet is inherently unstable, and cells maintain the monolayer integrity by regulating their density, topology, and apicobasal polarity.

 The state diagram indicates that the apical and basal extrusion states (A and B) became stabilized when $5 \ge n \ge 4$.
Under those conditions, the cells formed rosette-like structures (illustrated in Fig. \ref{fig:3}d). The rosette-like structures obtained in the model (Fig. \ref{fig:4}c) are also observed in extruding cells in several physiological contexts \cite{rosenblatt2001epithelial,marinari2012live}. Rosette structures have often been observed on the apical side of epithelial monolayers \cite{harding2014roles}; however, our model suggests that they can also form on the basal surface. The lack of prior observations of rosette-like structures on the basal surface might be due to the difficulty in observing cell-cell boundaries on the basal side of live epithelial monolayers.

\subsection*{Active generation of forces on cell-cell boundaries provokes extrusion}

\begin{figure}
\begin{center}
\includegraphics[scale=1.5]{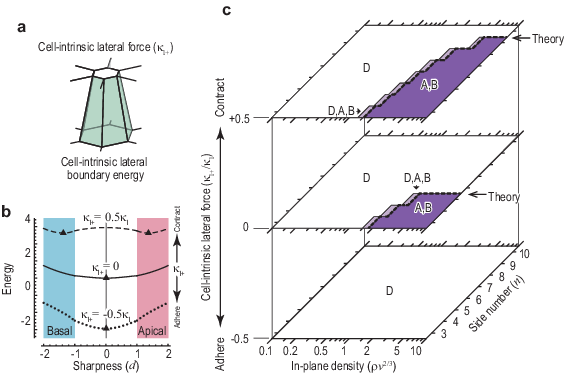}
\end{center}
\caption{
Effects of active forces exerted on lateral cell-cell boundaries of extruding cells.
{\bf a}. Force specifically generated on the lateral boundary areas between the center cell and its first neighbors. {\bf b}. Energy landscape of Eq. \ref{LateralEnergy} with respect to the active lateral force $\kappa_\text{l+}$. Triangles ($\blacktriangle$) indicate energy minimum points. {\bf c}. State diagram of Eq. \ref{LateralEnergy} with respect to the number of sides $n$, the in-plane cell density $\rho$, and the active lateral force $\kappa_\text{l+}$. The color coding in {\bf b} and {\bf c} is identical to that in Fig. \ref{fig:3}a and \ref{fig:3}d, respectively. The dashed lines indicate the theoretical transition boundaries between stable and unstable regions of 2 dual-sided state (D), as described by Eq. \ref{Criticaldensity2}. The physical parameters were set to $k_\text{a} = 0$, $n = 6$, and $\rho = 2.5 V^{-\frac{2}{3}}$ in a, and $k_\text{a} = 0$ in b.
}
\label{fig:4}
\end{figure}

Extruding cells occasionally generate active forces, making the mechanical properties heterogeneous within the cellular sheet. For example, actomyosin accumulates along the apicobasal axis and generates contractile force, as observed in invertebrate apoptosis \cite{monier2015apico}, morphogenesis \cite{sui2018differential}, and vertebrate cell extrusion \cite{kajita2014filamin}. Because mechanical instability is inherent in the force balance on cell-cell boundaries, the active generation of force on those boundaries might enhance the instability and provoke extrusion.

We assumed that single cells generate forces on the boundaries with their neighbors (Fig. \ref{fig:4}a). We described those forces as a lateral energy $\kappa_\text{l+} A_\text{c}$, proportional to the lateral boundary area between the center cell and its first neighbors $A_\text{c}$. The variable $\kappa_\text{l+}$ indicates the relative difference between the lateral forces exerted on $A_\text{c}$ and those exerted on $A_\text{p}$. Positive and negative $\kappa_\text{l+}$ values represent contractile and adhesive forces, respectively. Thus, Eq. \ref{TotalEnergy} can be rewritten as

\begin{equation}
\label{LateralEnergy}
U_\text{l} = 
  U_\text{p}
+ \kappa_\text{l+} A_\text{c}.
\end{equation}

To analyze the effects of  $\kappa_\text{l+}$ on cell extrusions, we focused on $\left| d \right| \ll 1$ and calculated the Mclaurin series of Eq. \ref{LateralEnergy} at $k_\text{a}=0$. Similarly to Eq. \ref{Criticaldensity}, the analytical calculation gave the critical density $\rho_\text{l+*}$, described as
\begin{equation}
\label{Criticaldensity2}
\rho_\text{l+*} = 2 \left( 
	\frac{ 3 \cos \left( \frac{\pi}{n} \right) }{
	n V^2 \left( 2 \sin \left( \frac{\pi}{n} \right) 
	- \frac{\kappa_\text{l}}{\kappa_\text{l} + \kappa_\text{l+}} \right)
	}
	\right)^\frac{1}{3},
\end{equation}
where the state at $d=0$ is stable when $\rho < \rho_\text{l+*}$ and unstable when $\rho > \rho_\text{l+*}$.
Under the condition with $k_\text{a} = 0$, because Eq. \ref{LateralEnergy} is an even function of $d$, state bifurcations occur in a symmetry-breaking manner at $\rho = \rho_\text{l+*}$, depending on $n$, $\rho$, $V$, and the force balance between $\kappa_\text{l}$ and $\kappa_\text{l+}$.

To analyze cell behaviors across a wide range of $d$, we calculated energy landscapes and state diagrams using Eq. \ref{LateralEnergy} (Fig. \ref{fig:4}). The energy landscapes demonstrated that, similarly to increases in $\rho$, the generation of lateral force induces instability leading to extrusion (Fig. \ref{fig:4}b). The dependencies of the stability of dual-sided state (D) on $\rho$, $n$, and
$\kappa_\text{l+}$ (Fig. \ref{fig:4}c) are quantitatively consistent with the theoretical boundaries of Eq. \ref{Criticaldensity2}. The state diagrams obtained from Eq. \ref{LateralEnergy} indicate
that lateral constriction ($\kappa_\text{l+}>0$) causes extrusion even when $n \ge 7$, whereas lateral adhesion ($\kappa_\text{l+}<0$) suppresses extrusion even when $n \le 6$ (Fig. \ref{fig:4}c). Thus, active lateral constriction lead to cell extrusion independently of the foam geometry of the epithelial sheet.




\subsection*{Asymmetrical force generation directs extrusion to the apical or basal side of the cellular sheet}

\begin{figure}
\begin{center}
\includegraphics[scale=1.5]{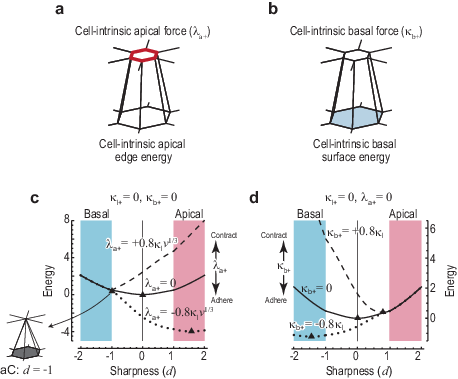}
\end{center}
\caption{
Individual effects of active forces exerted on the apical and basal sides of extruding cells.
{\bf a}. Active force specifically exerted on the apical perimeter of the center cell.
{\bf b}. Active force specifically exerted on the basal area of the center cell.
{\bf c}--{\bf d}. Energy landscapes of Eq. \ref{ApicobasalEnergy} with respect to apical $\lambda_\text{a+}$ and basal $\kappa_\text{b+}$ forces, respectively. 
Triangles ($\blacktriangle$) indicate energy minima.
Physical parameters were set to $k_\text{a}=0$, $n=6$, and $\rho=1.5 V^{-\frac{2}{3}}$.
}
\label{fig:5}
\end{figure}

We investigated the individual roles of forces generated on the apical and basal sides of an extruding cell. Under physiological conditions, the apical and basal regions of an extruding cell generate forces such as myosin-induced contraction and cadherin-mediated or integrin-mediated adhesion \cite{katoh2012epithelial,gudipaty2017epithelial,rosenblatt2001epithelial,monier2015apico,kajita2014filamin}. The geometric regions where the forces are exerted are asymmetric: the apical forces are exerted on the apical perimeter, whereas the basal forces are exerted on the whole basal surface. 

We considered the additional apical and basal energies separately by making the additional apical energy, $\lambda_\text{a+} P_\text{c}$, proportional to the apical junction length $P_\text{c}$ (Fig. \ref{fig:5}a) and the additional basal energy, $\kappa_\text{b+} A_\text{c}$, proportional to the basal surface area $A_\text{b}$ (Fig. \ref{fig:5}b).
Thus, Eq. \ref{LateralEnergy} could be rewritten as
\begin{equation}
\label{ApicobasalEnergy}
U_\text{a} = 
  U_\text{l}
+ \lambda_\text{a+} P_\text{c}
+ \kappa_\text{b+} A_\text{b}.
\end{equation}

To understand roles of the apical and basal forces cell extrusion, we calculated energy landscapes using Eq. \ref{ApicobasalEnergy}. The landscapes demonstrated that apical and basal adhesions ($\lambda_\text{a+}<0$, $\kappa_\text{b+}<0$) stabilized the apical and basal extrusion states (A and B), respectively (Fig. \ref{fig:5}c,d).
Basal constriction ($\kappa_\text{b+}>0$) maintained the dual-sided state (D), whereas apical constriction ($\lambda_\text{a+}>0$) stabilized the apical convergence state (aC), allowing the cells to stably maintain a pseudostratified structure. Those results indicate that the force regulation in apicobasal regions determines the direction of cell extrusion to either the apical side or the basal side of the cellular sheet.

\begin{figure}
\begin{center}
\includegraphics[scale=1.5]{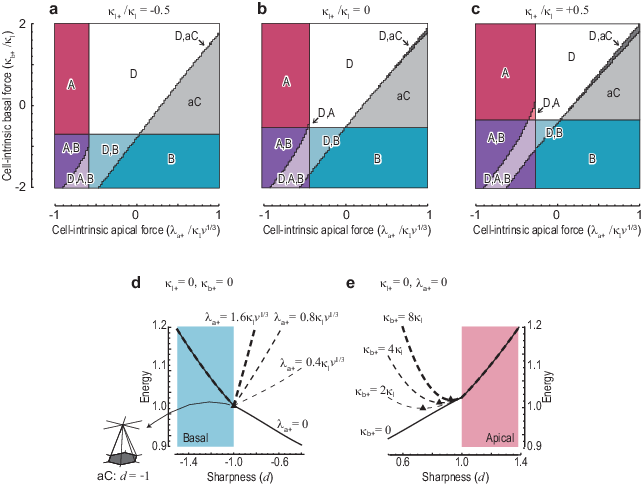}
\end{center}
\caption{
Combined effects of active forces exerted on the apical, basal, and lateral sides of extruding cells.
{\bf a}--{\bf c}. State diagrams of Eq. \ref{ApicobasalEnergy} with respect to apical $\lambda_\text{a+}$, basal $\kappa_\text{b+}$, and lateral $\kappa_\text{l+}$ forces. Physical parameters were set to $k_\text{a}=0$, $n=6$, and $\rho=1.5 V^{-\frac{2}{3}}$.
The cell states are color coded: white, unistable state (D); red, unistable state (A); blue unistable state (B); light gray, unistable state (aC); light red, bistable state (D, A); light blue, bistable state (D, B); dark gray, bistable state (D, aC); purple, bistable state (A, B); light purple, tristable state (D, A, B). The abbreviations of states (D), (A), (B), and (aC) are identical to those in  Fig. \ref{fig:1}e. 
{\bf d}--{\bf e}. Energy landscapes of Eq. \ref{ApicobasalEnergy} with respect to the apical energy $\lambda_\text{a+}$ and $\kappa_\text{b+}$, respectively. Triangles ($\blacktriangle$) indicate energy minima.
Physical parameters were set to $n = 6$, $\rho = 1.5 V^{-\frac{2}{3}}$, and $\kappa_\text{l+} = 0$.
}
\label{fig:6}
\end{figure}

Under physiological conditions, the active forces on the apical, lateral, and basal regions of the cells in epithelial sheets are often intertwined. For example, the expression of Rho family proteins that downregulate actomyosin activity is deteriorated in human tumors \cite{del2005rho,abraham2001motility}, upsetting the balance among the apical, basal, and lateral forces. To understand the combined effects of the separate active forces, we calculated state diagrams using Eq. \ref{ApicobasalEnergy}. The state diagrams demonstrated that the combination of active apical, basal, and lateral forces induces several bifurcations of the stable cell states [e.g., the bistable states (A, B), (D, A), (D, B), and (D, aC) in Fig. \ref{fig:6}]. Overall, as $\kappa_\text{l+}$ increased, the stable region of the dual-sided state (D) gradually shrunk (white region in Fig. \ref{fig:6}a--c). A decrease in either the apical $\lambda_\text{a+}$ force or the basal $\kappa_\text{b+}$ force directed extrusion selectively to the apical (A) or basal (B) side, respectively (Fig. \ref{fig:6}a--c).


The dependence of cell states on the apical ($\lambda_\text{a+}$) and basal ($\kappa_\text{b+}$) forces was asymmetric (Fig. \ref{fig:6}); that is, the diagrams involved the bistable state (D, aC) but not the bistable state (D, bC). The asymmetry emerges from the difference in the geometric regions where the apical and basal forces are exerted. The geometric difference leads to a difference in stable cell states via mechanical energy. In order to clarify the effects of the apical and basal forces on the mechanical instability of the apical and basal convergence states, we focused on the behaviors of Eq. \ref{ApicobasalEnergy} around $d = -1$ and $d = 1$. Because the apical and basal forces affect cell states under conditions where $d \ge -1$ and $d \le 1$, respectively, we narrowed the range of $d$ to $-1 \le d \le 1$. In addition, because Eq. \ref{ApicobasalEnergy} is discontinuous at $d=-1$ and $d=1$, we considered the behaviors around the limits  $d \to -1^{+}$ and $d \to 1^{-}$ under $k_\text{a}=0$ and $\kappa_\text{l+}=0$. In the absence of the basal force ($\kappa_\text{b+}=0$), the Taylor series of Eq. \ref{ApicobasalEnergy} around $d=-1^{+}$ is written as
\begin{equation}
\label{TaylorApical}
U_\text{a} = \begin{array}{ll}
	\kappa_\text{l} g_\text{0}
	+ \left( \kappa_\text{l} g_\text{1} + \lambda_\text{a+} \sqrt{ \cfrac{3 n \tan \left( \frac{\pi}{n} \right) }{\rho} } \right) \left( d + 1 \right) + O \left( \left[ d + 1 \right]^2 \right)
& \text{as $-1 < d < 1$}
\end{array}
.
\end{equation}
In Eq. \ref{TaylorApical}, $g_\text{0}$, $g_\text{1}$, and $g_\text{2}$ are the zeroth, first-order, and second-order coefficients of the lateral energy ($\kappa_\text{l}$) as functions of $n$, $\rho$, and $V$.
In the absence of the apical force ($\lambda_\text{a+}=0$), the Taylor series of Eq. \ref{ApicobasalEnergy} around $d=1^{-}$ is written as
\begin{equation}
\label{TaylorBasal}
U_\text{a} = \begin{array}{ll}
	\kappa_\text{l} h_\text{0}
	+ \kappa_\text{l} h_\text{1} \left( d - 1 \right)
	+ \left( \kappa_\text{l} h_\text{2} + \cfrac{3 \kappa_\text{b+}}{4 \rho} \right) \left( d - 1 \right)^2 + O \left( \left[ d - 1 \right]^3 \right)
& \text{as $-1 < d < 1$}
\end{array}
.
\end{equation}
In Eq. \ref{TaylorBasal}, $h_\text{0}$, $h_\text{1}$, and $h_\text{2}$ are the zeroth, first-order, and second-order coefficients of the lateral energy ($\kappa_\text{l}$) as functions of $n$, $\rho$, and $V$. 
Equations \ref{TaylorApical} and \ref{TaylorBasal} show the different dependences of the energy functions on $\lambda_\text{a+}$ and $\kappa_\text{b+}$. Because the lowest order of the apical energy ($\lambda_\text{a+}$) around $d = -1^{+}$ is linear and similar to that of $\kappa_\text{l}$, the stability of the apical convergence state ($d = -1$) strongly depends on $\lambda_\text{a+}$. On the other hand, because the lowest-order term of the basal energy ($\kappa_\text{b+}$) around $d = 1^{-}$ is higher than that of $\kappa_\text{l}$, the stability at the basal convergence state ($d = 1$) slightly depends on $\kappa_\text{b+}$. These results indicate that even when the convergence states are not stable under the influence of the lateral energy ($\kappa_\text{l}$), the apical energy ($\lambda_\text{a+}$) has a potential to stabilize the apical convergence state ($d = -1$), as observed in Fig. \ref{fig:6}. Therefore, the asymmetric localization of mechanical factors in the subcellular level leads to forming the asymmetric layer structures of epithelial sheets in the apicobasal axis.

\section*{DISCUSSION}
We described the discrete and 3D multicellular mechanics of epithelial sheets and analytically demonstrated that there is inherent instability in the 3D foam geometry of those structures that is sufficient to drive cell extrusion. Our results indicate that the maintenance of cellular monolayers requires a proper 3D cell geometry and force balance, which are ignored in most 2D models \cite{hannezo2014theory,honda1982cell,farhadifar2007the,misra2016shape}. Our results suggest that the stability of cellular monolayers depends on the cell geometry and the generation of various active forces under {\it in vivo} and {\it in vitro} physiological conditions, indicating a deep interdependence among the 3D foam geometry, cellular force generation, and multicellular integrity in epithelia. Furthermore, our results suggest that a mechanical instability is the fundamental mechanism that drives and determines the direction of cell extrusions. Our model also explains various epithelial geometries such as rosette structures and pseudostratified structures, suggesting that it might be applicable to other phenomena such as epithelial homeostasis \cite{macara2014epithelial} and the formation of layered structures \cite{miyata2001asymmetric,haubensak2004neurons,rackley2012building}.
The simple condition in Fig. \ref{fig:3} corresponds to the physiological situation where an extruding cell and its neighbors possess homogeneous mechanical properties. The resulting dependences on $k_\text{a}$, $n$, and $\rho$ agree well with experimental observations, suggesting that mechanical instability is a determinant of the 3D foam geometry of epithelial sheets. For example, epithelial sheets usually have a hexagonal packing geometry $n \approx 6$) and contain few cells with $n=3$ \cite{gibson2006emergence,farhadifar2007the}. Upon extrusion {\it in vivo}, most cells reduce their number of topological neighbors from $n \approx 6$ to $n \approx 3$ \cite{marinari2012live}. That dependence on $n$ under physiological conditions is qualitatively consistent with the dependence of the stability of the dual-sided state (D) on $n$ in our model (Fig. \ref{fig:3}b,d). Furthermore, under physiological conditions, extrusion is accompanied by an increase in cell density \cite{marinari2012live,eisenhoffer2012crowding,levayer2016tissue}, which is qualitatively consistent with the dependence of the stability of the dual-sided state (D) on $\rho$ in our model (Fig. \ref{fig:3}b,d). The parameter ranges of $n$ and $\rho$ surveyed in our study approximately correspond to those observed in various physiological contexts. The transition from the dual-sided state to the extrusion state under physiological conditions is accompanied by a loss of apicobasal cell polarity \cite{katoh2012epithelial,gudipaty2017epithelial}, which corresponds to a decrease in $k_\text{a}$ in our model (Fig. \ref{fig:3}d). The many agreements between our model and observations in living systems suggest that the relationship between mechanical instability and cell extrusions that exists in our model also exists in living systems.

The conditions modeled in Figs. \ref{fig:4}--\ref{fig:6} correspond to the situation where an extruding cell actively generates forces according to an intrinsic genetic program. The resulting dependences on $\kappa_\text{l+}$, $\lambda_\text{a+}$ correspond to a wide range of physiological conditions. For example, apoptotic cells accumulate actomyosin along the apicobasal (lateral) axis \cite{monier2015apico}, implying
that they actively cause lateral constriction (positive $\kappa_\text{l+}$) to help drive the extrusion process (Fig. \ref{fig:4}). Similar actomyosin accumulation is observed in several other physiological contexts \cite{sui2018differential,kajita2014filamin}. Apoptotic cells also form an actomyosin ring along the boundary with their neighbors \cite{katoh2012epithelial,gudipaty2017epithelial,rosenblatt2001epithelial,monier2015apico,kajita2014filamin}, implying that they actively cause constriction on either the apical side or the basal side (positive $\lambda_\text{a+}$ or positive $\kappa_\text{b+}$), thus determining the direction of extrusion (Figs. \ref{fig:5}--\ref{fig:6}). E-cadherin knockdown in normal cells that surround Ras-positive cells reduces the frequency of apical extrusion of the Ras-positive cells while promoting basal protrusion formation and invasion \cite{hogan2009characterization}. Because cadherin-mediated cell-cell adhesion is coupled to actin accumulation along adherens junctions, the knockdown of E-cadherin reduces the apical contractility of the normal cells surrounding the Ras-positive cells, which corresponds to the relative increase in the apical contractility of the center cell (i.e.,  $\lambda_\text{a+}$) in our model (Fig. \ref{fig:6}). The initial state of basal cell invasion, corresponding to the basal extrusion state (B) in our model, is accompanied by integrin-mediated adhesion \cite{friedl2003tumour}, which corresponds to a relative increase in the basal adhesion of the center cell (i.e., decrease in $\kappa_\text{b+}$) in our model (Fig. \ref{fig:6}). Thus, the effects of apical, lateral, and basal forces ($\lambda_\text{a+}$, $\kappa_\text{l+}$, and $\kappa_\text{b+}$ in Figs. \ref{fig:4}--\ref{fig:6}) in our model are consistent with local accumulations of mechanical factors such as actomyosin, cadherin, and integrin in living systems. Those results indicate that cells can exert forces within epithelial sheets in order to utilize the mechanical instability inherent in the 3D foam geometry of the monolayer. Therefore, the spatial distributions of the molecules that produce those forces should be rigorously regulated at the subcellular level. Furthermore, it is possible that defects in those distributions lead to pathologies related to epithelial integrity and extrusion.
  Although our model recapitulates physiological processes, it is limited by some approximations. 

First, the geometry of the cells in the model is simplified as a flat sheet, whereas epithelial sheets can be curved by
forces that produce invagination or branching \cite{gomez2018scutoids, kim2013apical}. Such curvatures might affect the mechanical stability of epithelial sheets. For example, a recent study suggested that the curvature of an epithelial sheet affects whether a given cell is extruded to the apical or the basal side \cite{maechler2019curvature}. Another recent study suggested that the cell geometry in curved sheets can vary more widely than that in flat sheets \cite{gomez2018scutoids}. Moreover, from a topological point of view, it is known that pentagon-like and heptagon-like defects accumulate in regions with positive and negative curvatures, respectively \cite{irvine2010pleats}. Although it is challenging to analyze the effects of curvatures on the mechanical stability of epithelial sheets, it might be possible to conduct such an analysis by regarding the sheet curvature as a constraint condition. A second approximation that limits our model is that individual cell shapes are simplified as polyhedrons with flat surfaces, whereas actual cell surfaces can be curved. The same simplification is employed in vertex models, which nevertheless have succeeded in recapitulating various developmental processes \cite{fletcher2014vertex}. We applied our model to qualitatively analyze cell extrusions under conditions representing various physiological conditions (Fig. \ref{fig:3}--\ref{fig:6}). More quantitative analyses might be made possible by taking into account changes in cell surface curvatures. A third limitation of our model is that the mechanical energies of the apical and basal surfaces are simply expressed as linear terms of the surface areas, and their contributions to the total energy in the flat cellular sheet are ignored. Although simple functions are often used in the modeling of mechanical cell behaviors \cite{honda2004a,okuda2013reversible,okuda2018strain}, higher-order terms might also affect the instability of cell sheet, which could be addressed by modifying the mechanical energy function. A fourth limitation of our model is that it focuses on steady states of single-cell behavior. Under physiological conditions, there are temporal changes in active force generation during the extrusion process \cite{hogan2009characterization,friedl2003tumour,kajita2014filamin}. Moreover, cell extrusions can sometimes be coupled to surrounding cell behaviors such as proliferation, apoptosis, and motility or migration. Such dynamic and multicellular behaviors can be modeled by computational simulations using 3D vertex models \cite{honda2004a,okuda2013reversible,okuda2018strain,messal2019tissue}, which have the ability to link the basic insights gained through our model to specific physiological contexts. Despite its limitations, our model provides a guide to understanding the wide range of epithelial physiology that occurs in morphogenesis, homeostasis, and carcinogenesis.


\section*{CONCLUSION}
We presented a theoretical model that describes the mechanical stability of cellular monolayers. The model shows that the cellular monolayer has inherent mechanical instability that is sufficient to cause cell extrusion without any additional provoking factors, even when the cells have homogenous mechanical properties. Analytical calculations show that the instability depends on the cell density, topology, and mechanical properties within the monolayer. The active generation of forces by an extruding cell can provoke extrusion and direct the extrusion to the apical or basal side of the monolayer, suggesting that cells utilize the mechanical instability of the epithelial geometry in order to regulate extrusion.

\section*{Author Contributions}

S.O. conceived of the study, carried out the calculations, and wrote the manuscript. K.F. provided critical feedback and helped to write the manuscript.
 
\section*{Acknowledgments}

We thank Dr. Tetsuya Hiraiwa at the University of Tokyo, Dr. Romain Levayer at Institut Pasteur, Dr. Yosuke Ogura at RIKEN, Dr. Katsuyoshi Matsushita at Osaka University, and Mr. Koki Nunota at Kanazawa University for discussions.
S.O. thanks his colleagues in the laboratory of Prof. Mototsugu Eiraku at Kyoto University for discussions.
This work was supported by the JST CREST Grant No. JPMJCR1921 (S.O.), JST PRESTO Grant No. JPMJPR16F3 (S.O.), JSPS KAKENHI Grant No. 17KT0021 (S.O.), 17H02939 (S.O.), 17H05619 (K.F.), and 19H04777 (S.O.), the Uehara Memorial Foundation (S.O.), and by the World Premier International Research Center Initiative (WPI), MEXT, Japan (S.O.).

\bibliography{reference}


\section*{Supplementary Material}

\subsection*{Derivation of the mechanical energy functions}

When the center cell shape is expressed as a regular frustum, it is uniquely determined by a set of $n$, $V$, $A_\text{a}$, and $A_\text{b}$ (Fig. \ref{fig:1}c). 
The volume of the frustum is calculated as
\begin{equation}
\label{frustum}
V = \frac{H \left( A_\text{a} + \sqrt{A_\text{a} A_\text{b}} + A_\text{b} \right)}{3}.
\end{equation}
The topology of the center cell depending on $d$ (Fig. \ref{fig:1}e) gives
\begin{equation}
\label{topology}
\left\{
\begin{array}{lllll}
A_\text{a}=0 & \text{when} & d \le -1 \\
H=T & \text{when} & -1 \le d \le 1 \\
A_\text{b}=0 & \text{when} & d \ge 1 \\
\end{array}
\right..
\end{equation}

We simultaneously solved Eqs. \ref{sharpness}, \ref{frustum}, \ref{topology} and obtained $H$, $A_\text{a}$ and $A_\text{b}$ as functions of $T$, $V$, and $d$. By replacing $T$ with $\rho$ using $\rho=T/V$, we obtained $H$, $A_\text{a}$ and $A_\text{b}$ as functions of $\rho$, $V$, and $d$ as follows:
\begin{equation}
\label{cellheight}
H \left( \rho, V, d \right) = \begin{cases}
	- \cfrac{\rho V}{d} & 
	d \le -1 \\
	\rho V  & 
	-1 < d < 1 \\
	\cfrac{\rho V}{d} & 
	d \ge 1
\end{cases},
\end{equation}
\begin{equation}
\label{apicalarea}
A_\text{a} \left( \rho, d \right) = \begin{cases}
	0 & 
	d \le -1 \\
	\cfrac{ 3 \left( d + 1 \right)^2 }{\rho \left( d^2 + 3 \right) } & 
	-1 < d < 1 \\
	- \cfrac{ 3 d }{ \rho } & d \ge 1
\end{cases},
\end{equation}
\begin{equation}
\label{basalarea}
A_\text{b} \left( \rho, d \right) = \begin{cases}
	- \cfrac{ 3 d }{ \rho } & 
	d \le -1 \\
	\cfrac{ 3 \left( d - 1 \right)^2 }{\rho \left( d^2 + 3 \right) } & 
	-1 < d < 1 \\
	0 & 
	d \ge 1
\end{cases}.
\end{equation}

\begin{figure}
\begin{center}
\includegraphics[scale=1.5]{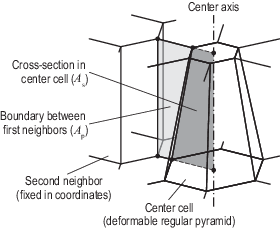}
\end{center}
\caption{
3D foam geometry of the center, first-neighbor, and second neighbor cells in a planar sheet.
The center cell shape is represented by a deformable regular frustum. The center axis is defined as the axis of the center cell aligned along the normal direction to the sheet. The boundary faces between the first neighbor cells are aligned radially from the center axis. These boundary faces connect to the second neighbor cells, whose geometry is fixed in position. The cross-sections are defined as the regions bounded by the center axis and the lateral edge of the center cell (colored dark gray). Total area of $n$ cross-sections is represented by $A_\text{s}$. The boundary faces between the first-neighbor cells are defined as the regions bounded by the edge of the center cell and the lateral edge of the second-neighbor cells (colored light gray). The total area of $n$ boundary faces is represented by $A_\text{p}$. Because the center axis and the second-neighbor cells are fixed in position, the total area of $A_\text{s}$ + $A_\text{p}$ ($=C$) is constant. Therefore, $A_\text{p}$ is derived as $C - A_\text{s}$.
}
\label{fig:7}
\end{figure}

The geometric parameters $P_\text{c}$, $A_\text{c}$, and $A_\text{p}$ in the energy function $U$ are described as functions of $n$, $H$, $A_\text{a}$, and $A_\text{b}$ as follows:
\begin{equation}
\label{apicalperimeter}
P_\text{c} \left( n, A_\text{a} \right) = 
	2 \sqrt{ n A_\text{a} \tan \left( \frac{\pi}{n} \right)},
\end{equation}
\begin{equation}
\label{lateralarea}
A_\text{c} \left( n, H, A_\text{a}, A_\text{b} \right) = 
	\left( \sqrt{A_\text{a}} + \sqrt{A_\text{b}} \right)
	\sqrt{ A_\text{a} - 2 \sqrt{ A_\text{a} A_\text{b} } + A_\text{b} + H^2 n \tan \left( \frac{\pi}{n} \right)},
\end{equation}
\begin{equation}
\label{surroundingarea}
A_\text{p} \left( n, H, A_\text{a}, A_\text{b} \right) = 
	C - H \sqrt{ \frac{ n }{ 2 \sin \left( \frac{2 \pi}{n} \right) } } \left( \sqrt{A_\text{a}} + \sqrt{A_\text{b}} \right),
\end{equation}
where $C$ is a constant that represents the total area of the boundary faces between neighbor cells and the cross-sections in the center cell (Fig. \ref{fig:7}). 
Because $A_\text{p}$ is incorporated into the energy function in Eq. \ref{TotalEnergy} as the first order (i.e., $\kappa_\text{l} A_\text{p}$), $C$ is negligible. Therefore, we set $C = 0$ for simplification. By substituting Eqs. \ref{cellheight}--\ref{basalarea} into Eqs. \ref{apicalperimeter}--\ref{surroundingarea}, we obtained $P_\text{c}$, $A_\text{c}$, and $A_\text{p}$ as functions of $n$, $\rho$, $V$, and $d$ as follows:
\begin{equation}
\label{apicalperimeter2}
P_\text{c} \left( n, \rho, d \right) = \begin{cases}
	0 & 
	d \le -1 \\
	2 \sqrt{ \cfrac{3 n \left( d + 1 \right)^2 \tan \left( \frac{\pi}{n} \right) }{ \rho \left( d^2 + 3 \right) } } &
	-1 < d < 1 \\
	2 \sqrt{ \cfrac{3 n d \tan \left( \frac{\pi}{n} \right) }{ \rho } } &
	d \ge 1
\end{cases},
\end{equation}
\begin{equation}
\label{lateralarea2}
A_\text{c} \left( n, \rho, V, d \right) = \begin{cases}
	\sqrt{ \cfrac{9d^2}{\rho^2} - \cfrac{3 n \rho V^2 \tan \left( \frac{\pi}{n} \right)}{d} } & 
	d \le -1 \\
	\cfrac{2}{\rho \left( d^2 + 3 \right)} \sqrt{ 36 d^2 + 3 \left( d^2 + 3 \right) n \rho^3 V^2 \tan \left( \frac{\pi}{n} \right) } &
	-1 < d < 1 \\
	\sqrt{ \cfrac{9d^2}{\rho^2} + \cfrac{3 n \rho V^2 \tan \left( \frac{\pi}{n} \right)}{d} } & 
	d \ge 1
\end{cases},
\end{equation}
\begin{equation}
\label{surroundingarea2}
A_\text{p} \left( n, \rho, V, d \right) = \begin{cases}
	- \sqrt{ \cfrac{ - 3 n V^2 \rho }{ 2 d \sin \left( \frac{2 \pi}{n} \right) }} &
	d \le -1 \\
	- \sqrt{ \cfrac{ 6 n V^2 \rho }{ \left( d^2 + 3 \right) \sin \left( \frac{2 \pi}{n} \right) }} &
	-1 < d < 1 \\
	- \sqrt{ \cfrac{ 3 n V^2 \rho }{ 2 d \sin \left( \frac{2 \pi}{n} \right) }} &
	d \ge 1 \\
\end{cases}.
\end{equation}
Therefore, by substituting Eqs. \ref{apicalarea}, \ref{basalarea}, \ref{apicalperimeter2}--\ref{surroundingarea2} into Eqs. \ref{LocalEnergy}, \ref{LateralEnergy}, \ref{ApicobasalEnergy}, we obtained analytical expressions of the individual energy functions.

\subsection*{Derivation of the local apical perimeter energy}

Assuming that more cells than just the first-neighbor cells are fixed in position, we considered the positions of the center and
first-neighbor cells. The perimeter of the first-neighbor cells is represented by $P_\text{p}$. Therefore, the contribution of the apical perimeter energy is given as
\begin{equation}
\label{apicalperimeterenergy}
U_\text{a} = k_\text{a} \left( P_\text{c} - P_\text{eq} \right)^2 + n k_\text{a} \left( P_\text{p} - P_\text{eq} \right)^2.
\end{equation}
For simplification, we assume that $P_\text{p} = P_\text{eq}$ when $P_\text{c} = P_\text{eq}$. By calculating the change in $P_\text{p}$ by the change in $P_\text{c}$, $P_\text{p}$ is described as a function of $P_\text{c}$, $P_\text{eq}$, and $n$, by
\begin{equation}
\label{surroundingperimeter}
P_\text{p} = P_\text{eq} + \cfrac{ 1 - \csc \left( \frac{\pi}{n} \right) }{n} \left( P_\text{c} - P_\text{eq} \right),
\end{equation}
where the coefficient of the second term $\left( 1 - \csc \left[ \pi / n \right] \right) / n$ indicates the ratio of the apical perimeter strain of the first neighbor cell to that of the center cell.
By substituting Eq. \ref{surroundingperimeter} into Eq. \ref{apicalperimeterenergy}, we obtained the term of the apical perimeter energy in Eq. \ref{LocalEnergy}.
Notably, the results (Fig. \ref{fig:3}b--c, a part of Fig. \ref{fig:3}d, Fig. \ref{fig:4}--\ref{fig:6}) are independent of the approximation used in the derivation of the apical perimeter energy, because $k_\text{a}$ is set to zero.

\subsection*{Analytical calculation of the critical density}

By assuming the condition with $\left| d \right| \ll 1$ and $k_\text{a}=0$, we obtained the McLaurin series of Eq. \ref{LocalEnergy} for $d$ as Eq. \ref{McLaurin}. The expression of $f_\text{2}$ in Eq. \ref{McLaurin} is given as
\begin{equation}
\label{McLaurin2}
f_\text{2} \left( n, \rho, V \right) = 
	\frac{1}{3 n \rho^4 V} \left(  
	12 \sqrt{ n \rho^3 \cot \left[ \frac{\pi}{n} \right] }
	- V^2 \sqrt{ n^3 \rho^9 \tan \left[ \frac{\pi}{n} \right] }
	\right)
	+ \frac{V \csc \left[ \frac{\pi}{n} \right]}{6} \sqrt{ n \rho \tan \left[ \frac{\pi}{n} \right] }
,
\end{equation}
where the first and second terms are associated with $A_\text{c}$ and $A_\text{p}$, respectively. By solving $f_\text{2}=0$ using Eq. \ref{McLaurin2}, we obtained the critical density described by Eq. \ref{Criticaldensity}.
The critical density of Eq. \ref{Criticaldensity2} is obtained in the same way.


\end{document}